# Magnetic Ordering of Ammonium Cations in $NH_4I$, $NH_4Br$ and $NH_4Cl$

Fei Yen[1,2*], Lei Meng[1], Tian Gao[3], Sixia Hu[4]

[1]School of Science, Harbin Institute of Technology, Shenzhen, University Town, Shenzhen, Guangdong 518055 P. R. China
[2]State Key Laboratory on Tunable Laser Technology, Ministry of Industry and Information Technology Key Laboratory of Micro-Nano Optoelectronic Information System, Harbin Institute of Technology, Shenzhen, University Town, Shenzhen, Guangdong 518055, P. R. China
[3]School of Mathematics and Physics, Shanghai University of Electric Power, 2588 Changyang Road, Shanghai 200090, P. R. China
[4]Sustech Core Research Facilities, Southern University of Science and Technology, 1088 Xueyuan Blvd. Shenzhen, Guangdong 518055, P. R. China

**Correspondence:** *fyen@hit.edu.cn, +86-1343-058-9183;

**Abstract:** The different types of magnetism arise mainly from how electrons move and interact with each other. In this work, we show how protons ($H^+$) also exhibit magnetic behavior. We measured the magnetic susceptibility of the ammonium halides and identified pronounced increases at 232 K, 233 K and 243 K for $NH_4I$, $NH_4Br$ and $NH_4Cl$, respectively, which all coincide to the geometric ordering of its ammonium cations. With extensive literature establishing the fact that the ammonium cations exhibit rotational motion even towards the lowest temperatures, we take into account that the orbital motion of the protons carries a magnetic moment and find it to be larger than that of the paired electrons. Consequently, the structural phase transitions are magnetically-driven as the system attempts to lift 8-fold energy degeneracies of the proton orbitals via Jahn-Teller distortions. Our findings identify that $NH_4^+$ cations are capable of comprising magnetism which appears to be ubiquitous in ammonia-based molecular solids.

**Introduction:**

The methyl groups $CH_3$ in hexamethylbenzene $C_6(CH_3)_6$ were recently found to order magnetically below 118 K.[1,2] As each methyl group rotates about its own axis, its three protons ($H^+$) at the extremities generate current loops which upon multiplying it by the encircling area gives rise to an orbital magnetic moment $\mu_p$ along the axial direction.[3] The electrons of the hydrogen atoms are treated on average to reside midway the C-H bonds comprising part of a singlet. Hence, the distance travelled by a proton per revolution of a methyl group rotation is twice than its two electrons forming the C-H bond. This results in $\mu_p$ being twice as large (and opposite in direction) as the magnetic moment of the two C-H bond electrons because $\mu = I\pi r^2$ (where $I$ is the current and $r$ is the radius). Analogous are the π-electrons in aromatic compounds; these electrons cover a much larger area when acting as induced currents in the benzene rings when under an applied field resulting in a more 'exalted' diamagnetism.[4,5] Moreover, the intrinsic spins of the protons in each methyl group may be treated to be unpaired according to Hund's rule resulting in the overall magnetic moment of the protons as being much larger than the diamagnetic contributions of the remaining paired electrons. It may therefore be interpreted that each rotating methyl group possesses three 'valance protons' upon which below a finite temperature, long range order should be established. Naturally, we wondered whether similar magnetic behavior occurs in ammonium ($NH_4^+$) cations which are not only slightly more massive than methyl groups but also possess a larger rotational constant and a tetrahedron geometry.[6] We selected the ammonium halides ($NH_4I$, $NH_4Br$, $NH_4Cl$) as the $NH_4^+$ are well known to exhibit rotational tunneling even toward low temperatures.[7-18] Nearly every sort of study has been performed on these ammonium salts,[19-26] except their magnetic properties since the susceptibility of all diamagnets is expected to remain negative and independent all the way down to absolute zero,[27] and to our knowledge, no researchers have taken into account the notion that rotating $NH_4^+$ carry a residual magnetic moment after partitioning the positive and negative charge components.

The high temperature phase of the ammonium halides crystallizes into the NaCl structure,[28] space group $Fm3m$, denoted as the α phase (or phase I) in existing literature. The $NH_4^+$ cations and $X^-$ anions (where X=I, Br and Cl) take the Na and Cl sites, respectively. At lower temperature, the system transitions into the β phase (II) which has the CsCl structure,[20] space group $Pm3m$, where each of the $NH_4^+$ and $X^-$ are situated at the center of a cube surrounded by the other species at the corners. The four protons ($H^+$) lie along the center line of the N and X atoms in a tetrahedral arrangement so there are two possible $NH_4^+$ orientations (Fig. 1a) *A* and *B* which are randomly distributed (disordered). Table 1 lists the critical temperatures between the different phases. For $NH_4I$ and $NH_4Br$, there exists a γ phase (III) at an even lower temperature (space group $P4/nmm$).[20,29] It possesses a tetragonal-like structure with the $X^-$ anions situated slightly off-centered by 0.012 Å and the $NH_4^+$ orientations become geometrically ordered in the sense that each adjacent unit cell along the *a* and *b* axes possess alternating *A* and *B* configurations reminiscent of a chessboard. The *A* and *B* configurations form columns along the *c* axis. For $NH_4Br$ and $NH_4Cl$, a δ phase (IV) exists at the lowest temperatures at ambient pressure (space group $P:3m$) which also has a CsCl-like structure[30] but the orientations of all of the $NH_4^+$ align parallel to each other,[20] *viz.* either all *A* or all *B*. The $NH_4^+$ tetrahedra undergo $C_3$ and $C_2$ rotations which are essentially simultaneous 'jumps' of three and four protons by 120° and 180°, respectively, to each other's sites (grey arrows in Fig. 1a). $C_4$ translations by 90° also occur, but this entails a change of configuration from *A* to *B* or vice versa. The $C_3$ rotations are more energetically favorable so they become more dominant with cooling.[8-10,18,31] Akin to the methyl groups in hexamethylbenzene, the $NH_4^+$ cations also possess a net magnetic moment upon experiencing a $C_3$ rotation if we treat the electrons of the hydrogen atoms to also lie near midway the N-H bonds. This is because the area of the orbits of the electrons and protons are different so their respective magnetic moments do not cancel out. Figures 1b and 1c provide a graphical representation of this model showing how the electron and proton components add up to yield a net magnetic moment. In the α and β phases the orientation distribution of the $NH_4^+$ tetrahedra is random so the net magnetization contribution from all of the individual $\mu_p$ is zero. We hypothesized that in the γ and δ phases, it is the magnetic interactions between $NH_4^+$ moments that causes their anti-parallel and parallel ordering. As such, the $NH_4^+$ moments in these two ordered phases should also become susceptible to applied magnetic fields and subjected to canting to yield a contribution to the net magnetization which should be detectable by a magnetometer. This prompted us to perform magnetic susceptibility measurements on polycrystalline $NH_4I$, $NH_4Br$ and $NH_4Cl$ in the temperature range of 2-300 K and external magnetic fields of up to 50 kOe presented below.

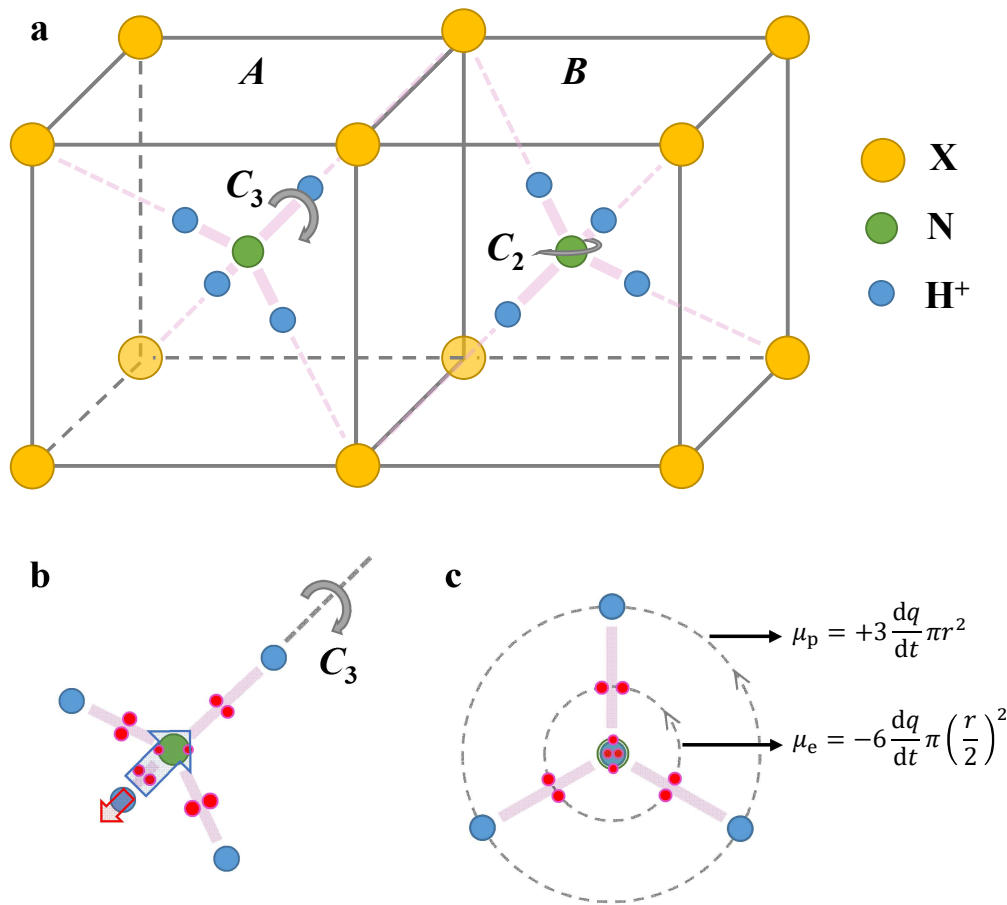

**Figure 1:** a) The β phase of ammonium halide $NH_4X$ where X=I, Br or Cl. The $C_3$ and $C_2$ rotations are when the $NH_4^+$ cations rotate about a diagonal or a principle axis, respectively. Two possible configurations *A* and *B* exist according to how the $NH_4^+$ is oriented in the lattice which is random in the β phase. b) Side view of the $NH_4^+$ cation. Red spheres represent average positions of the electrons. When experiencing a $C_3$ rotation, the axis of rotation comprised by the N atom and one H atom along with the two electrons bonding these two atoms and the two s-shell electrons of the N atom remain 'stationary'. As the remaining 3 protons and 6 electrons orbit about the axis of rotation, two orbital magnetic moment components (indicated by the blue and red

arrows) are generated due to the opposite nature of their charges. c) The same $NH_4^+$ cation projected along the axis of rotation, *i.e.* along one of the vertices of the tetrahedron. The orbital moment μ may be taken as the product of the current $I = dq/dt$ and the area encircled by the orbit of the protons. It is clear that the orbital moment of the protons $\mu_p$ is twice as large as that of the electrons' $\mu_e$ if the electrons are treated on average to reside midway the N and H atoms. However, it is expected that $\mu_p \gg \mu_e$ since the averaged positions of the electrons reside closer to the N atoms due to it being 7 times more positively charged than the H atoms.

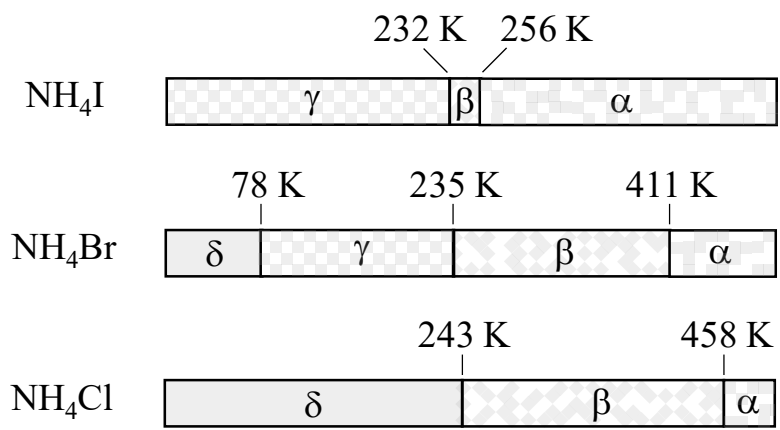

**Table 1:** The critical temperatures of the structural phase transitions of the ammonium halides at ambient pressure with α and β as disordered and γ and δ as ordered phases. The small white and grey squares represent configurations *A* and *B*.

**Methods:**

Polycrystalline samples of ammonium iodide NH4I (99.99% in purity) and ammonium chloride NH4Cl (99.99% in purity) were acquired from Aladdin, Inc. (Shanghai, P. R. China). Polycrystalline samples of ammonium bromide NH4Br (99.99% in purity) were acquired from Sigma Aldrich, Inc. For the NH4I sample, a Magnetic Properties Measurement System by Quantum Design, Inc. (U.S.A.) was employed to measure the magnetic susceptibility via its vibration sample magnetometer (VSM) option; for the field dependent measurements, the DC option was used. For NH4Br and NH4Cl, a Physical Properties Measurements System (PPMS-Dynacool) by Quantum Design, Inc. (U.S.A.) with its VSM option was used to measure the magnetic susceptibility. The temperature sweeping speeds were always maintained between 1 and 3 K/min and the applied magnetic fields were ramped by no more than 100 Oe/s. The size of the samples was usually around 1 mm$^3$; they were attached onto a quartz palette by GE Varnish and allowed to cure for 5 minutes prior to measurements. The crystal lattices were drawn and analyzed by the freeware VESTA (Visualization for Electronic and STructural Analysis) version 3.4.6.[32]

**Results:**

Figure 2 shows the magnetic susceptibility with respect to temperature $\chi(T)$ of NH$_4$I under applied magnetic fields of $H$=5 and 10 kOe. During cooling, a sharp step discontinuity was observed at $T_{\beta\text{-}\gamma\_I}$=232 K. Toward lower temperatures, a shoulder-type of anomaly is identified near 40 K and starting from below 13 K $\chi(T)$ started to diverge and behave as $T^{-0.5}$. The dashed line is the supposed behavior of $\chi(T)$ if the system was strictly diamagnetic, that is, if only the electrons contributed to the magnetization. During warming, the shoulder-type anomaly appears to have morphed into a peak and shifted to near 45 K. Near 280 K, a step-down discontinuity was also observed. Figure 3 shows $\chi(T)$ of NH$_4$Br under $H$=50 kOe as this curve has the least noise; more $\chi(T)$ curves at lower $H$ showing similar features are provided in the Supporting Information section. A much less pronounced discontinuity was observed at $T_{\beta\text{-}\gamma\_Br}$=233 K during cooling. However, $\chi(T)$ decreased by over 20% starting from $T_{\beta\text{-}\gamma\_Br}$ down to 10 K; this is in contrast to only 10% in NH$_4$I going from $T_{\beta\text{-}\gamma\_I}$ to 10 K. A maximum was also identified in NH$_4$Br near 55 K along with a change of slope discontinuity at 177 K upon warming. Figure 4 displays $\chi(T)$ of NH$_4$Cl under $H$=10 kOe where the slope suddenly changed from near zero to becoming negative at $T_{\beta\text{-}\delta\_Cl}$=135 K during cooling and the reverse occurred at $T_{\delta\text{-}\beta\_Cl}$=243 K during warming. Similar to NH$_4$Br, a maximum was observed at 75 K and a change of slope discontinuity at 175 K in NH$_4$Cl.

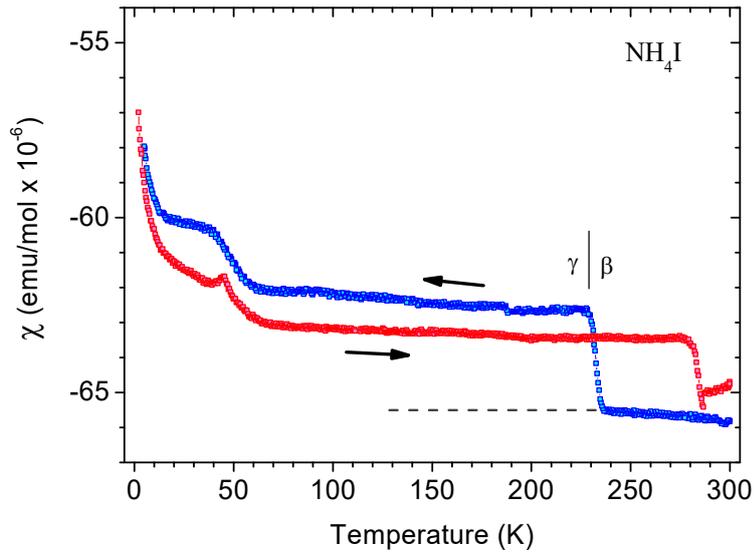

**Figure 2:** Magnetic susceptibility χ(*T*) with respect to temperature of ammonium iodide NH4I under an applied magnetic field of *H*=5 kOe during cooling and 10 kOe during warming. Dashed line represents χ(*T*) if NH4I was purely diamagnetic. The short vertical line demarcates the β to γ phase transition at $T_{β-γ\_I}$=232 K.

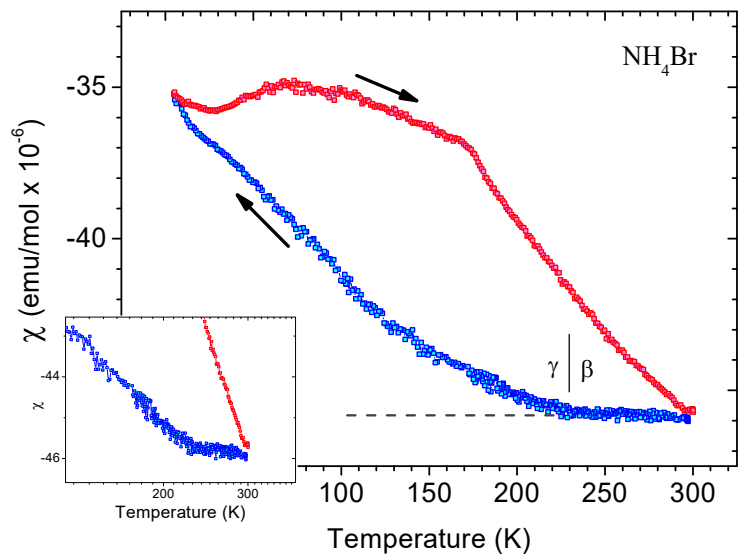

**Figure 3:** χ(*T*) of ammonium bromide NH4Br with *H*=50 kOe. Inset is an enlargement (in logarithmic scale) near the β to γ phase transition occurring at $T_{β-γ\_Br}$=233 K.

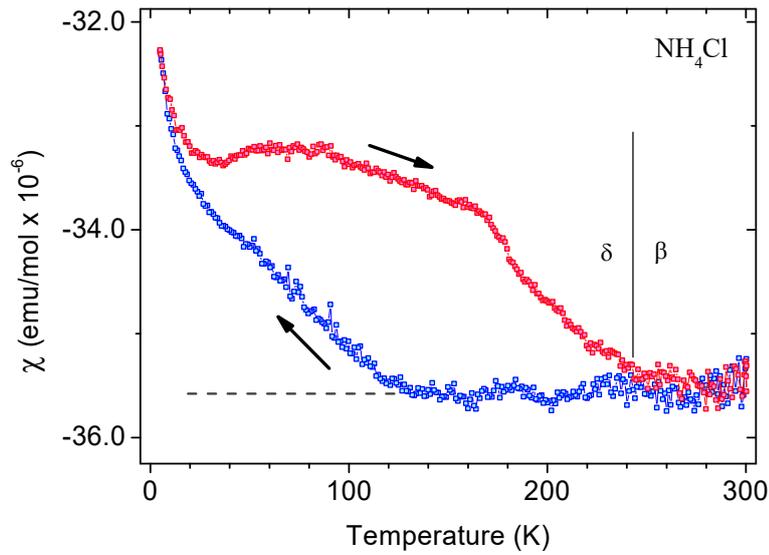

**Figure 4:** χ(*T*) of NH₄Cl at 10 kOe. Vertical line shows where the system phase transitioned from δ to β at $T_{\delta\text{-}\beta\_Cl}$=243 K.

Figure 5 shows the magnetization under varying applied magnetic fields *M(H)* held at constant temperatures for the three studied ammonium halides. The *M(H)* curves for NH₄I and NH₄Cl (Figs. 5a and 5b) remained linear down to the lowest measured temperatures, however, with a systematic decrease in slope. For the case of NH₄Br, starting from 25 K, *M(H)* became nonlinear and became positive in slope when *H*<1.5 kOe (Fig. 5c).

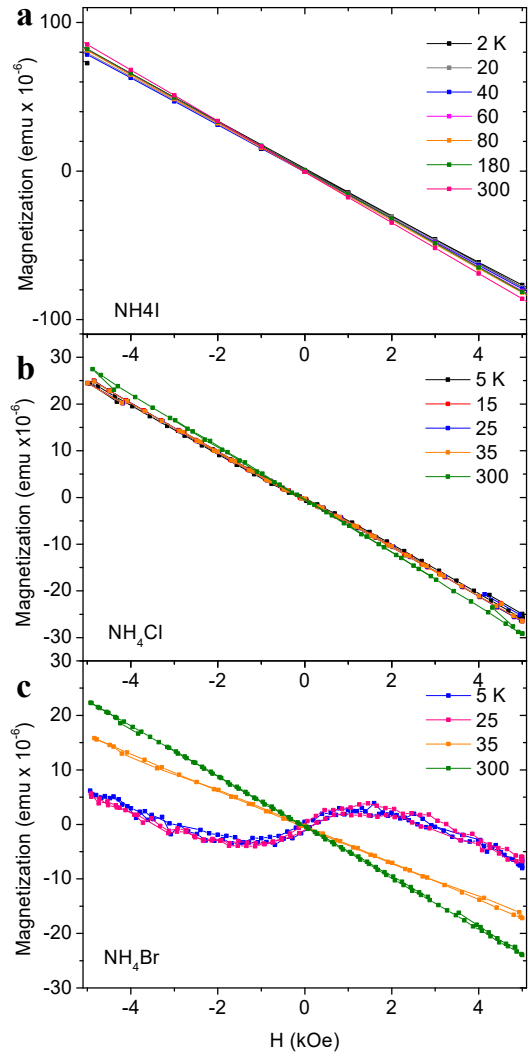

**Figure 5:** Magnetization in arbitrary units versus $H$ for a) NH$_4$I, b) NH$_4$Cl and c) NH$_4$Br at constant temperatures.

**Discussion:**

The increase of $\chi(T)$ at $T_{\beta\text{-}\gamma\_I}$, $T_{\beta\text{-}\gamma\_Br}$ and $T_{\delta\text{-}\beta\_Cl}$ all coincide to the temperatures at which the disordered NH$_4^+$ become either ordered anti-parallel or parallel with respect to each other. In contrast, no anomalies were observed in $\chi(T)$ in all of the other structural phase transitions between disordered to disordered (such as α to β at 255.8 K in NH$_4$I) and ordered to ordered (γ to δ at 78 K in NH$_4$Br) phases. These two results are near irrefutable evidence that the orientational ordering of the NH$_4^+$ cations are magnetic in nature. To check if this is theoretically possible, we calculated the spin and orbital magnetic moments of each proton of an $NH_4^+$ cation experiencing $C_3$ rotations which have torsional frequencies ranging from 279 to 397 cm$^{-1}$ according to specific heat,[33] inelastic neutron scattering, Raman spectroscopy and NMR measurements summarized in Ref. [34]; details are found in the Supporting Information section. The obtained effective magnetic moment is 0.0269 $\mu_B$. This figure is rather small compared to ordinary antiferromagnetic systems possessing similar values above unity. However, all electrons are paired and the distance between two $NH_4^+$ is only 4 Å apart. Even more important is that the radii at which the protons orbit is definite and is more than twice than that of unpaired electrons. These two factors lead to proton-to-proton distances between adjacent $NH_4^+$ of around 2.87 Å which make it possible for the $NH_4^+$ to order antiferromagnetically. Lastly, the anomalies associated to the ordering of the $NH_4^+$ is most pronounced in $NH_4I$ followed by $NH_4Br$ and $NH_4Cl$; this may be explained by the results of Gutowsky et al. as well as Leung et al. concluding that the potential barriers to rotation is lowest in $NH_4I$ and highest in $NH_4Cl$.[10,34]

The ammonium halides are known for their ability to be supercooled and superheated by large temperature ranges.[16,25,30,33,35-38] This explains the large hysteretic regions observed in the warming and cooling $\chi(T)$ curves. From such, the increase at $T_{\beta-\delta\_Cl}$=135 K in $\chi(T)$ (Fig. 5) was due to the β phase of $NH_4Cl$ remaining supercooled all the way down to 135 K before transitioning into the δ phase. By the same token, the anomaly at $T_{\gamma-\beta\_Br}$=295 K in $\chi(T)$ of Fig. 3 was due to the γ phase of $NH_4Br$ persisting all the way up to 295 K before transitioning back to the β phase. Observation of martensitic behavior and presence of mixed phases of β and γ or β and δ has also been reported.[35,36] Hence, the maxima and change of slope anomalies below the order-disorder transitions during warming in $NH_4Br$ and $NH_4Cl$ may be due to a phase transitioning of a mixed phase or domain wall motion. In the low temperature regime, the $T^{-0.5}$ dependence of $M(T)$ below 13 K for $NH_4I$ and $NH_4Cl$ appear to not be due to impurities, otherwise the dependence would be $T^{-1}$ according to Curie's Law. Our best explanation to this behavior is that the stationary proton in each $NH_4^+$ (the one not involved in translational motion unlike the other three in a $C_3$ rotation) which also has a spin 1/2 paramagnetic contribution but has zero orbital angular momentum, is strongly correlated to its $NH_4^+$ ensemble so a $T^{-1}$ dependence is no longer possible. The current approach explaining the orbital and spin component contributions is too crude a model. Any proton spin-orbit coupling and electron-proton coupling should also be considered as they may be what underlie the pronounced features and sharp increases of $\chi(T)$ toward low temperatures.

The $M(H)$ curves do not exhibit hysteretic regions which further verify that the discrepancy between the $\chi(T)$ cooling and warming curves in the $\gamma$ and $\delta$ phases are not due to ferromagnetism. The decreasing of the $M(H)$ slopes with lowering temperature suggests that the magnetic interactions between $NH_4^+$ intensify correspondingly. In the special case of $NH_4Br$, the slope of $M(H)$ becomes positive below 25 K and 1.5 kOe indicative of the magnetization component stemming from the $NH_4^+$ cations as becoming more dominant than the diamagnetism generated by the electrons. However, we note that a presence of magnetic impurities in $NH_4Br$ where the magnetic moments become saturated at 1.5 kOe is also possible and cannot be ruled out. An absence of a hysteretic region in the $M(H)$ curves in the $\delta$ phases of $NH_4Br$ and $NH_4Cl$ suggest that despite all adjacent $NH_4^+$ tetrahedra are aligned geometrically parallel to each other as suggested by neutron scattering experiments,[20] their magnetic moments align antiferromagnetically similar as in the $\gamma$ phase.

There are eight possible ways for the $NH_4^+$ tetrahedra to perform a $C_3$ rotation (clockwise and counterclockwise along each of its four vertices). This renders the protons to possess an 8-fold orbital degeneracy so when the $NH_4^+$ cations begin to interact magnetically the system must lift this degeneracy according to the Jahn-Teller effect.[39] One way to reason this is to treat the protons in each rotating $NH_4^+$ to also possess a natural frequency, one that is not oscillatory between two sites but instead about three sites periodically and sequentially when experiencing $C_3$ rotations. This means that adjacent $NH_4^+$ cations resonate at each others' natural frequencies which create an interacting force. When thermal noise is low enough to no longer disrupt these intermolecular forces, geometric distortions appear. According to neutron diffraction measurements, the Br⁻ anions are oppositely displaced by 0.012 Å in the γ phase[20] and numerous reports have concluded the δ phase as being piezoelectric.[40,41] Hence, the distortion of the halide framework which aligns adjacent $NH_4^+$ cations to become either parallel or anti-parallel to each other is magnetically-driven. This new perspective is in great contrast to two widely employed models in describing the transitions between the β and γ (or δ for the case of $NH_4Cl$) phases. The first model is that of Nagamiya's where the order-disorder transition depends on the coherency of the electrostatic potential of the $NH_4^+$ cations.[42] The second theory by Yamada et al. employs an Ising model to represent the two possible tetrahedron configurations and it was found that the two ordered phases are results of direct $NH_4^+$ interactions and indirect $NH_4^+$ interactions mediated by phonons.[43] We emphasize that this theory did not assign a magnetic moment component to the rotating $NH_4^+$. Our model is not only capable of explaining the observed magnetic properties reported herein, but can also pinpoint the λ-peak in the heat capacity as arising from magnetic entropy due to the breaking of time-reversal symmetry from the ordering of the proton orbital magnetic moments as well as why the ordered phases have uniquely distorted lattices which occur not only in the ammonium halides but also in most other molecular solids. The idea that rotating protons also carry a magnetic moment may also be applied to higher ordered moieties that undergo rotation; one extreme example being that of rotating boron cages in $(NH_4)_2B_{10}H_{10}$ and $(NH_4)_2B_{12}H_{12}$.[44] Even if the involved protons only experience torsional librations, if there is energy degeneracy of their motion, then resonant forces may also trigger a phase transition. Such may be the case in ice, however, the cohesive forces binding the lattice may be too large so a full transition is suppressed leading to the existence of metastable phases.[45-47]

**Conclusions:**

To conclude, we identify the presence of inter-molecular magnetic interactions in the ammonium halides $NH_4I$, $NH_4Br$ and $BH_4Cl$ through temperature and external magnetic field dependent measurements of the magnetic susceptibility. The onset temperatures of a non-diamagnetic component coincide to the respective critical temperatures where the $NH_4^+$ tetrahedra become structurally ordered lending unequivocal evidence that $NH_4^+$ cations, and not electrons, are responsible for the observed magnetism. Due to the 8-fold energy degeneracy of the $C_3$ rotations, when the $NH_4^+$ moments attempt to order, Jahn-Teller distortions lift the energy degeneracies which trigger the β to γ (β to δ) structural phase transitions in $NH_4I$ and $NH_4Br$ ($NH_4Cl$). The resulting unique configuration of the magnetic moments resembles that of a Type-G antiferromagnetic system, albeit their spin directions are aligned along the diagonals rather than along a principle axis because of the geometry of the proton orbitals. The rotational motion of the $NH_4^+$ (more specifically, the orbital motion of protons) in the ammonium halides appear to play a crucial role on the stability of the system as they govern the structural, thermal and magnetic properties. Apart from this work, there are thousands of ammonium-based compounds whose magnetic properties at low temperatures remain to be investigated. Of particular interest are those that are ferroelectric such as ammonium sulfate $(NH_4)_2SO_4$ which to this day, it is not yet clear why it becomes ferroelectric at $T_c$=223 K.[48-51] Our preliminary results on seed crystals of $(NH_4)_2SO_4$ reveal magnetic anomalies present at $T_c$ along with magnetic anisotropy. By taking into account that the $NH_4^+$ tetrahedra (of which there exists two different sites in $(NH_4)_2SO_4$) possess magnetic moments due to orbital motion of its protons, the ordered magnetic structure in the low temperature phase is identified to break spatial-inversion symmetry suggesting that the spontaneous polarization is magnetically-driven. Our findings serve as a stepping stone to exploring a new breed of multiferroic compounds.

**Supporting Information:**

Figures S1, S2, S3 and a detailed description of the calculation of the magnetic moment and orbital angular momentum of protons of an $NH_4^+$ cation undergoing $C_3$ rotations.

**Acknowledgements:**

The authors are indebted to the National Natural Science Foundation of China for their financial support via The Research Fund for International Young Scientists and the Shanghai Municipal Natural Science Foundation grant number 16ZR1413600.

**Table of Contents Graphic:**

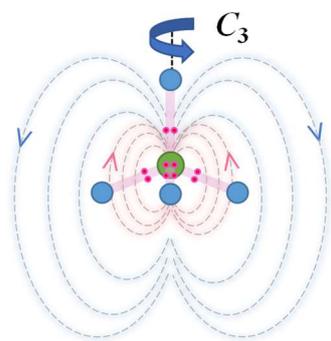

# Magnetic Ordering of Ammonium Cations in NH₄I, NH₄Br and NH₄Cl

Fei Yen[1,2*], Lei Meng[1], Tian Gao[3], Sixia Hu[4]

[1]School of Science, Harbin Institute of Technology, Shenzhen, University Town, Shenzhen, Guangdong 518055 P. R. China
[2]State Key Laboratory on Tunable Laser Technology, Ministry of Industry and Information Technology Key Laboratory of Micro-Nano Optoelectronic Information System, Harbin Institute of Technology, Shenzhen, University Town, Shenzhen, Guangdong 518055, P. R. China
[3]School of Mathematics and Physics, Shanghai University of Electric Power, 2588 Changyang Road, Shanghai 200090, P. R. China
[4]Sustech Core Research Facilities, Southern University of Science and Technology, 1088 Xueyuan Blvd. Shenzhen, Guangdong 518055, P. R. China

**Correspondence:** *fyen@hit.edu.cn, +86-1343-058-9183;

**Supporting Information:**

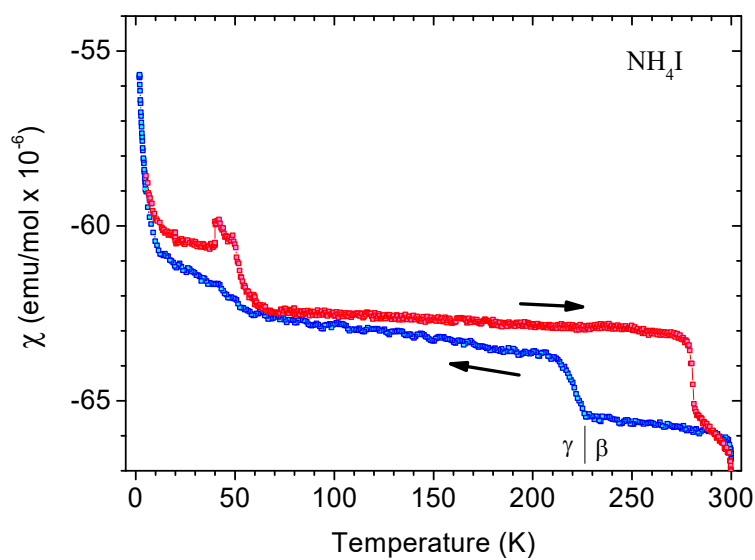

**Figure S1:** Additional data on the magnetic susceptibility χ(*T*) of NH₄I under applied magnetic fields of *H*=1 kOe during cooling and 20 kOe during warming. Vertical line demarcates the phase transition from β to γ at 228 K.

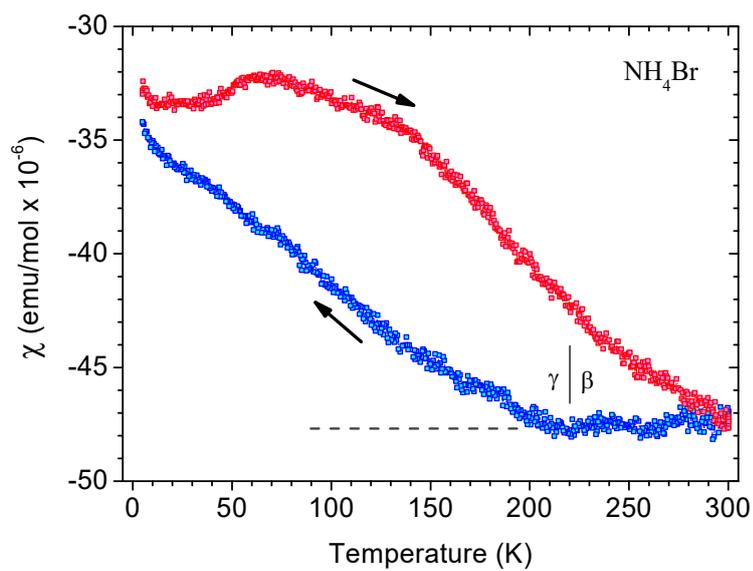

**Figure S2:** $\chi(T)$ of NH4Br at $H$=2 kOe. The phase transition from β to γ occurred at 221 K.

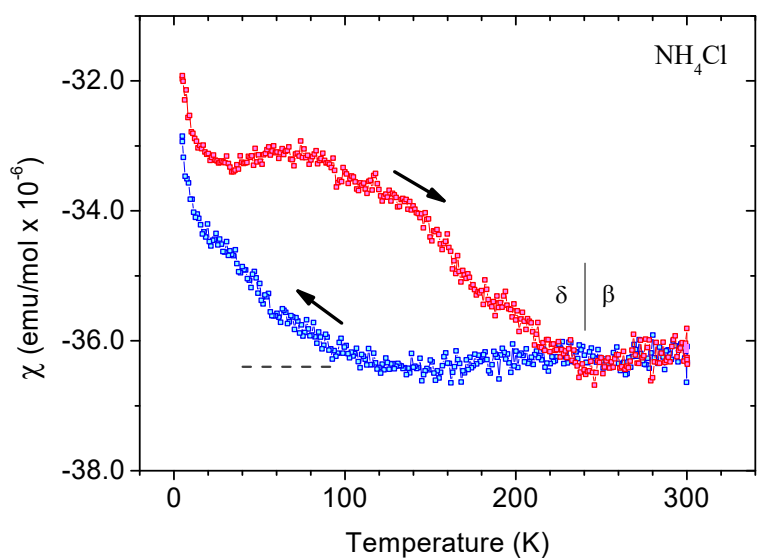

**Figure S3:** $\chi(T)$ of NH4Cl at $H$=2 kOe. The phase transition from δ to β during warming occurred at 239 K.

**The magnetic moment of an $NH_4^+$ cation undergoing $C_3$ rotations in the ammonium halides.**

From the experimentally obtained torsional frequencies of the $NH_4^+$ in ammonium halides, the orbital magnetic moment is first calculated. Then, together with the spin magnetic moment component, we obtain an effective magnetic moment.

The orbital magnetic moment of a proton in an $NH_4^+$ cation undergoing $C_3$ rotation may be calculated by the equation $\mu_L = I \cdot A = e \cdot f \cdot A$, where $I$, $A$, $f$ and $e$ are the current, enclosed area, frequency and the elemental unit of charge ($e = 1.602 \times 10^{-19}$ A·s), respectively. The torsional frequencies of the $NH_4^+$ range from 279 cm$^{-1}$ in $NH_4I$ to 397 cm$^{-1}$ in $NH_4Cl$.[33,34] With 1 cm$^{-1}$ = $1.986 \times 10^{-23}$ J, the corresponding energy of 279 cm$^{-1}$ is $K = 5.541 \times 10^{-21}$ J. The speed of the protons is then $v = 2,574$ m/s by using the equation $K = 0.5 m_p v^2$, where $m_p$ is the mass of the proton ($1.673 \times 10^{-27}$ kg). The frequency, obtained from $f = v / (2\pi \cdot r)$, where $r$ is the radius at which the protons rotate (in this case 0.731 Å, the distance between the rotation axis and each of the three orbiting protons), is $5.604 \times 10^{12}$ Hz. Hence, $\mu_L = e \cdot f \cdot A = 1.507 \times 10^{-26}$ A·m$^2$, or 0.0016 $\mu_B$ (where 1 $\mu_B = 9.274 \times 10^{-24}$ A·m$^2$ is the Bohr magneton). This value is rather small; however, the spin component is significantly larger as will be shown below. Note that this is the exact method to obtaining the value of 1 $\mu_B$ for the case of the electron in a hydrogen atom by presuming the speed of the electron as traveling 1/137 (the fine structure constant) of the speed of light while 'orbiting' about the Bohr radius, namely the Bohr theory. The obtained values for both the proton and electron cases are tabulated in Table S1 for ease of comparison.

On a side note, the magnitude of the orbital angular momentum of the proton $L = m_p v \times r = 3.148 \times 10^{-34}$ J·s, or $\approx 3\ \hbar$ (where $\hbar = 1.055 \times 10^{-34}$ J·s is the reduced Planck constant). This integer value is coincidental since for a torsional frequency of 397 cm$^{-1}$, the orbital angular momentum is $3.755 \times 10^{-34}$ J·s, or $\approx 3.5\ \hbar$. These elevated values from 1 $\hbar$ are due to the presence of periodic potential barriers.

We now calculate the spin magnetic moment component $\mu_S = g_S \cdot \mu \cdot S / \hbar$, where $g_S = 5.586$, $\mu = 1.411 \times 10^{-26}$ A·m$^2$ (2.793 times the nuclear magneton $\mu_N = 5.051 \times 10^{-27}$ A·m$^2$) and $S = 1/2\ \hbar$ are, respectively, the g-factor, intrinsic magnetic moment and the spin of the proton. From such values, we obtain $\mu_S = 3.941 \times 10^{-26}$ A·m$^2$, or 0.0042 $\mu_B$. Despite that only three protons experience orbital motion in $C_3$ rotations, the fourth proton not undergoing orbital motion should also possess an intrinsic spin so all four protons in $NH_4^+$ must possess a spin component and be unpaired according to Lenz's law. Hence, $\mu_{S\_Total} = 1.576 \times 10^{-25}$ A·m$^2$, or 0.017 $\mu_B$.

The nuclear spin contribution is similar but with a g-factor of 0.404, $S = 1\ \hbar$ and $\mu_N = 5.05 \times 10^{-27}$ A·m$^2$ for the nitrogen atom amounting to an extremely small value of $2.04 \times 10^{-27}$ A·m$^2$, or 0.00022 $\mu_B$. The spin magnetic moments of the electrons cancel each other out since they are all paired. The orbital magnetic moments of the six electrons forming the three bonds to the three protons experiencing orbital motion are also small since the orbits of the former have much smaller radii compared to that of the latter. From such, the total effective magnetic moment of $NH_4^+$ experiencing $C_3$ rotations at energies of 279 cm$^{-1}$ amounts to $\mu_J = 0.0219\ \mu_B$. This figure is still rather small compared to electrons ordering antiferromagnetically which usually possess effective moments larger than unity. However, the distances between the sites of unpaired electrons are usually much larger than 4 Å, which is the average distance between $NH_4^+$ sites. More importantly, the radii at which the protons orbit is more than twice than that of the mean radius of electrons (Bohr radius). Moreover, the electrons usually reside in regions smaller than the Bohr radius, whereas the proton to proton distance between two adjacent $NH_4^+$ is consistently only ~2.87 Å apart! It is these factors we believe that make it possible for the $NH_4^+$ in ammonium halides to order antiferromagnetically.

| | Proton in $NH_4^+$ | Electron in H atom | Comments |
|---|---|---|---|
| Energy, $K$ | 279 cm$^{-1}$ | 13.6 eV | 279 cm$^{-1}$ for $NH_4I$ and 397 cm$^{-1}$ for $NH_4Cl$ from Ref. 34 |
| ($K$ in Joules) | 5.541 x 10$^{-21}$ J | 2.179 x 10$^{-18}$ J | 1 cm$^{-1}$ = 1.986 x 10$^{-23}$ J, 1 eV = 1.602 x 10$^{-19}$ J |
| Mass, $m$ | 1.673 x 10$^{-27}$ kg | 9.109 x 10$^{-31}$ kg | $K = 0.5\ mv^2$ |
| Traveling velocity, $v$ | 2,574 m/s | $c/137$ | $c$ = 3 x 10$^8$ m/s, 1/137 is the fine structure constant |
| Frequency, $f$ | 5.604 x 10$^{12}$ Hz | 6.578 x 10$^{15}$ Hz | $f = v/2\pi \cdot r$ |
| Orbital radius, $r$ | 0.731 x 10$^{-10}$ m | $a_0$ = 5.292 x 10$^{-11}$ m | $a_0$ is the Bohr radius |
| Encircling area, $A$ | 1.679 x 10$^{-20}$ m$^2$ | 8.798 x 10$^{-21}$ m$^2$ | $A = \pi \cdot r^2$ |
| Orbital magnetic moment, $\mu_L$ | 1.507 x 10$^{-26}$ A·m$^2$ (0.0016 $\mu_B$) | 9.271 x 10$^{-24}$ A·m$^2$ (1 $\mu_B$) | $\mu_L = e \cdot f \cdot A$, $\mu_B$ is the Bohr magneton |
| Angular momentum, $L$ | 3.148 x 10$^{-34}$ J·s, $\approx 3\ \hbar$ | 1.055 x 10$^{-34}$ J·s, $\approx 1\ \hbar$ | $L = mv \times r$, $\hbar$ = 1.055 x 10$^{-34}$ J·s |
| Spin magnetic moment, $\mu_S$ | 3.941 x 10$^{-26}$ A·m$^2$ (0.0042 $\mu_B$) | 1 $\mu_B$ | $\mu_S = g_S \cdot \mu \cdot S/\hbar$, $\mu = \mu_B$ for electron, $\mu = 2.793 \mu_N$ for proton |
| Total magnetic moment, $\mu_J$ | 2.029 x 10$^{-25}$ A·m$^2$ 0.0219 $\mu_B$ | 2 $\mu_B$ | For $NH_4^+$, $\mu_J = 3 \cdot \mu_L + 4 \cdot \mu_S$ |

**Table S1:** The magnetic moment and orbital angular momentum calculated from the energy of a proton orbiting about an N atom in the $NH_4^+$ configuration undergoing $C_3$ rotations contrasted with the example of an electron orbiting about a proton in the hydrogen atom configuration; $e$ = 1.602 x 10$^{-19}$ A·s.

--